\documentclass{ws-ijmpa}

\newcommand{\dzero}     {D\O\  }

\begin{document}

\markboth{Reinhard Schwienhorst}
{The \dzero Run~II Trigger System}

%
\catchline{}{}{}{}{}
%

\title{The \dzero Run~II Trigger System}

\author{\footnotesize Reinhard Schwienhorst\footnote{
On behalf of the \dzero collaboration.}}

\address{Department of Physics and Astronomy, \\
Michigan State University, \\
East Lansing, Michigan 48824, USA. \\
        Email: schwier@fnal.gov}

\maketitle

\pub{Received (Day Month Year)}{Revised (Day Month Year)}

\begin{abstract}
I present the current status of the \dzero trigger system 
in Run~II at the Tevatron.

\end{abstract}

\section{Introduction}

The \dzero detector at the Fermilab Tevatron was upgraded for Run~II\cite{d0runI,d0upgrade}. This
upgrade included improvements to the trigger system in order to be able to handle the
increased Tevatron luminosity and higher bunch crossing rates compared to Run~I.

The \dzero Run~II trigger is a highly flexible system to select events to be written to tape
from an initial interaction rate of about 2.5~MHz. 
This is done in a three-tier pipelined, buffered system. The first tier (level~1) processes 
fast detector pick-off signals in a hardware/firmware based system to reduce the event rate to about 1.~5kHz. 
The second tier (level~2) uses information from level~1 and forms 
simple Physics objects to reduce the rate to about 850~Hz. The third tier (level~3) uses full detector 
readout and event reconstruction on a filter farm to reduce the rate to 20-30~Hz.
The \dzero trigger menu contains a wide variety of triggers. While the emphasis is on
triggering on generic lepton and jet final states, there are also trigger terms for specific 
final state signatures.

In this document we describe the \dzero trigger system as it was implemented and is currently
operating in Run~II.

\section{Level~1 and Level~2 Trigger System}
The first two levels of the trigger system reduce the event rate from the initial interaction
rate of about 2.5~MHz to a rate of below 850~Hz that allows for a full readout of the \dzero detector. 
In order to accomplish this, the first two levels are coupled together tightly through a system that coordinates 
the trigger decisions and distributes trigger information to the various sub-detectors. 
This ``trigger framework'' handles the information from the level~1 muon, calorimeter, and tracking
sub-systems as well as the level~2 trigger system. It forms global level~1 and level~2 trigger decisions
which are sent out to all detector systems to coordinate event transfers from front-end buffers to
level~1 buffers and to the level~3 system. 

\begin{figure}
\centerline{\psfig{file=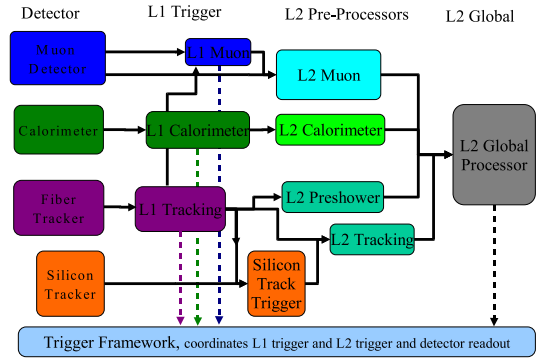,width=7cm}}
\vspace*{8pt}
\caption{A schematic illustration of \dzero Trigger level~1 and level~2 systems.}
\label{l1l2overview}
\end{figure}

A diagram of the level~1 and level~2 trigger configuration is shown in Fig.~\ref{l1l2overview}. At level~1,
the individual sub-systems are mostly independent, except for the ability to match muons to central tracks. 
At level~2, sub-detector specific objects are reconstructed in separate pre-processors. The level~2 global
processor then reads these objects from the pre-processors and combines them to form Physics objects. 
It furthermore computes event-wide variables such as the total transverse energy $H_T$ and 
event correlations such as $\phi$ separation between objects.

The level~1 and level~2 trigger systems allow for up to 128 individual triggers to be programmed. If any of these 
triggers passes, the event is read out and sent to level~3. Each individual trigger
may itself contain several different conditions. All conditions must be fulfilled in order for this trigger to
pass. This flexible configuration has enabled
the design of Physics trigger menus suited well for the instantaneous luminosities produced by the Tevatron
to date. During the Tevatron start-up, the focus of the trigger menu 
was on simple calorimeter-based objects (electrons and jets) and on muons reconstructed using the muon detector. 
As the instantaneous luminosity increased over time, track requirements, more complex objects, and event-wide 
variables were added.

\section{Level~3 Trigger System}
\begin{figure}
\centerline{\psfig{file=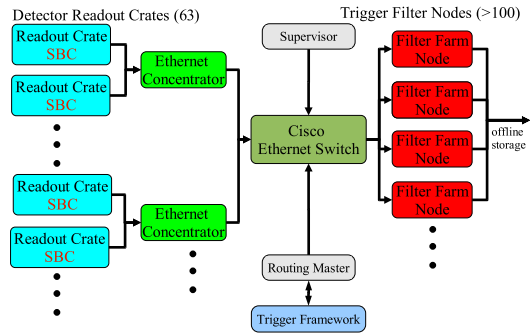,width=7cm}}
\vspace*{8pt}
\caption{A schematic illustration of \dzero Trigger level~3 system.}
\label{l3overview}
\end{figure}

The entire \dzero detector is read out for events passing level~1 and level~2. 
This includes reading out all of the detector elements as well as the trigger level~1 and level~2 systems themselves. 
Fig.~\ref{l3overview} shows an overview of the level~3 system. 

The hit and pulseheight information from each detector readout crate is collected in single-board-computers (SBC),
which send this information to a node on the filter farm through a commercial Ethernet switch. 
The flow of information from the readout crates to the filter nodes is controlled 
by the routing master. The trigger programming is loaded onto the filter nodes by the supervisor node. The
routing and filter node programming occurs over the same Ethernet links that are used in the data transfer. 

Each event is fully reconstructed at level~3 with algorithms that are similar to those used in the offline event 
reconstruction. This allows the level~3 system to accomplish a large rejection factor of 20 that is required to limit
the output rate to less than 50~Hz.
 
The flexibility of the trigger programming is expanded further at level~3. There are 256 individual trigger bits 
available at level~3, each coupled to one of the level~1/level~2 triggers. Each
can be programmed to filter on combinations of simple objects such as electrons, muons, or jets, as 
well as event-wide variables and correlations. Level~3 also provides the ability to select b-tagged jets based on 
tracking and silicon detector information.

\section{Conclusions}
The \dzero Run~II trigger system is working well, selecting events for offline processing with high efficiency
for many different Physics signal processes. Full advantage has been taken of the flexibility of the system as
the Tevatron instantaneous luminosity has increased.

\end{document}